\documentclass[11pt,english]{article}
\usepackage{ae,aecompl}
\usepackage{helvet}
\usepackage[T1]{fontenc}
\usepackage[numbers]{natbib}
\usepackage{amssymb,amsmath,amsthm}
\usepackage{graphicx}
\usepackage[pdfborder={0 0 0}]{hyperref}
\usepackage{chngpage}
\usepackage{pdflscape}
\usepackage{epstopdf}
\usepackage{fullpage}
\usepackage{setspace}
\usepackage{booktabs}
\usepackage[small,compact]{titlesec}
\usepackage{rotate}
\usepackage{subcaption}
\usepackage{float}
\usepackage{caption}
\usepackage{wrapfig}
\usepackage{sidecap}

\usepackage{hyperref}

\usepackage{multirow}
\usepackage{array}
\usepackage{graphicx}

\theoremstyle{plain} 
\theoremstyle{plain} 
\theoremstyle{plain}

\begin{document} 
	
	\title{Pandemic Policymaking: Learning the Lower Dimensional Manifold of Congressional Responsiveness}
	
	\author{Philip D. Waggoner \\ University of Chicago \\ \texttt{pdwaggoner@uchicago.edu}}
	
	\date{ }
	
	\maketitle
	
	\thispagestyle{empty} 

	\begin{abstract}
		\noindent A recent study leveraging text of pandemic-related policymaking from 1973--2020 explored whether pandemic policymaking has evolved, or whether we are witnessing a new, unique era of policymaking as it relates to large-scale crises like COVID-19. This research picks up on this approach over the same period of study and based on the same data, but excluding text. Instead, using high dimensional manifold learning, this study explores the latent structure of the pandemic policymaking space based only on bill-level characteristics. Results indicate the COVID-19 era of policymaking maps extremely closely onto prior periods of related policymaking. This suggests that there is less of an ``evolutionary trend'' in pandemic policymaking, where instead there is striking uniformity in Congressional policymaking related to these types of large-scale crises, despite being in a unique era of hyperpolarization, division, and ineffective governance. 
	\end{abstract}

\vspace{0.3in}


\vspace{0.7in}

\begin{center}
	\textbf{This is a working paper. Please contact the author prior to use or citation.}
\end{center}

	\clearpage
	\pagenumbering{arabic} 

\doublespacing

\section{Introduction}

COVID-19 has rattled the world with far reaching consequences from social \cite{alon2020impact} and political \cite{druckman2020political}, to epidemiological \cite{park2020systematic} and emotional \cite{lima2020emotional}. Though unprecedented on a number of dimensions, such large-scale societal crises require formalized responses to protect the vulnerable and needy in society. In the American context, when epidemics like COVID-19 occur, the clearest avenue for action that affects the broadest slice of the population is governmental action. The government, empowered by the constitution as well as the representational responsibility of political elites to respond to constituents \cite{eulau1977puzzle}, has the duty to act for the people by protecting, providing, and serving, especially when the country is facing a common threat, which currently is COVID-19. Governmental action in general can take many forms, such as court rulings, resource provision to workers on the ``front lines,'' military intervention, and support to those who need it most. While manifestations of governmental action vary, the most common avenue through which such large-scale governmental action takes shape, or is possible, is through policymaking. That is, writing legislation and passing bills into law reflect the government's prime means of action, distribution of resources, and mobilization of a unified response.\footnote{This is most simply seen through the Constitutional provision for Congress to ``hold the purse strings'' as it relates to monetary appropriations for governmental action \cite{ar1}. This monetary control and disbursement takes shape through policymaking, e.g., passing the federal budget bill.} In sum, policymaking by elected officials in Congress is one of the clearest avenues for governmental action, and especially in times of crisis.

COVID-19 is not the first large-scale crisis in America requiring a consolidated governmental response. Indeed many crises, both public health-related and otherwise, have littered America's past, from HIV/AIDS and the opioid epidemic, to urban crime, climate change, and public education. Yet, though the subject of the crisis might vary, there are several common threads underlying crises: 1) they affect a large portion of the population in some way (e.g., they are not geographically isolated), 2) the effects are negative in some measurable way, and 3) they require governmental policymaking to offer solutions, resources, and support to mitigate their effects. The result is these large-scale crises rise to the level of an ``epidemic'' or ``pandemic,'' depending on the scope of the issue.\footnote{Use of these terms is discussed at length below in the \textit{Empirical Strategy} section.}

Yet, the current climate of policymaking is different today than it was even a few decades ago. Though government need and responsibility have remained unchanged, the context and institution allowing for action taken through policymaking has drastically changed in several ways. Of note, the major political parties are extremely polarized and ideologically distinct \cite{mccarty2016polarized, pierson2020madison}; the country, comprised of a voting electorate, is deeply politically fractured \cite{abramowitz2005can}; driven largely by political activism, the alignment between the two major parties (Democrat and Republican) and the two major political ideologies (liberal and conservative) is stronger than ever before in American history \cite{noel2014political}; enactment of major rules changes has altered fundamental institutional processes, deepening elite political division \cite{kane2013reid}; and Congressional policymaking is increasingly negative and intense \cite{waggoner2020computational}. These recent and disheartening developments suggest America is experiencing a unique political period, at both the elite \textit{and} mass levels. 

In light of the COVID-19 pandemic, the relatively common occurrence of epidemics throughout American history, and the current era of hyper-polarization and division, a question emerges: \textit{is policymaking in response to COVID-19, which is occurring in this divided era, substantively different from policymaking in response to similar epidemics in the past?} This question assumes the occurrence of epidemics as well as the need for governmental response through policymaking are both constant. The question, then, centers on the \textit{nature} of policymaking and how it is currently taking shape, compared to recent history. There are a couple of ways one could approach addressing such a question: \textit{causal}, requiring theoretical innovation and development of testable expectations; or \textit{exploratory}, where nothing is assumed of the drivers of patterns in the data, seeking instead to learn from natural structure underlying the data. 

In a recent paper, \cite{waggoner2020computational} addressed a similar question using an exploratory framework, \textit{has governmental policymaking in response to epidemics and pandemics evolved, or are we witnessing a unique period of policymaking in the era of COVID-19?} To explore this question, \cite{waggoner2020computational} built and mined a set of the text data comprised of long bill titles, which can be dozens of words in length, and thus act as summaries and signals of the bill sponsor's intent. Results suggest that while the topics of epidemic-related bills historically remained focused on epidemics at respective moments in time (and are thus not ``evolutionary''), the general sentiment, or the ``how'' of the bills has substantially evolved, growing in both positive and negative sentiment over time. The increasingly intense tone defining the policymaking approach on these consequential, yet apolitical issues suggests the current COVID-19 era is indeed a distinct period of policymaking. Intensity and extremism are used to characterize proposed legislation in the current COVID-19 period, which is a marked shift from historical approaches to policymaking on similar issues. 

Yet a limitation of \cite{waggoner2020computational} is in the unit of analysis. While bill titles are useful mechanisms for ``bill branding,'' they are nonetheless a biased look at policymaking. In addition to numerous players contributing to crafting legislation (e.g., party leadership, multiple prime sponsors, and special interest groups), legislators themselves are inherently biased actors who have unique bases of support \cite{grossmann2016asymmetric}, political agendas to realize while in office \cite{grimmer2010bayesian}, a party brand to maintain and support \cite{butler2014understanding}, party leadership to satisfy \cite{ansolabehere2001effects}, and unique career concerns \cite{jenkins2005parties}. Further, legislators are ranked and rated by many special interest groups based on their policy portfolios \cite{poole1981dimensions, snyder1992artificial}, suggesting they may be biased \cite{kollman1997inviting} or at least influenced to write a bill that does not purely address the epidemic or issue in question, while instead posturing to obtain a better rating. In short, while the bill title is a useful summary and signal of the policy, and thus a good starting place to explore the historical contours of pandemic policymaking, this approach remains a tainted signal of a broader pattern of Congressional policymaking.

As such, this research picks up on this limitation, and is similarly motivated on the basis of the representational arrangement between policymakers and their constituents \cite{eulau1977puzzle, waggoner2019constituents}. Indeed, citizens of a representative democracy rightly expect policymakers who have been elected to serve an electing population to respond to large-scale issues like COVID-19. And accordingly, one of the most common avenues for \textit{responsiveness} by Congressional lawmakers is proposing and passing legislation, however tainted it may be, to address a problem from a variety of angles. 

Building on the growing body of research on COVID-19 and policymaking \cite{cairney2020covid, hartley2020policymaking, roy2020factors}, and specifically the evolutionary question of pandemic policymaking explored in \cite{waggoner2020computational}, I approach this question from a different angle, but using the same data set as \cite{waggoner2020computational}. I am interested in whether \textit{patterns} of policymaking are distinct or not. Put differently, are characteristics of bills addressing epidemics throughout recent history similar or different in meaningful ways? ``Meaningful ways'' in this context refers to structural similarity between the pre-COVID and COVID periods, e.g., similar or different cosponsor configurations and committee paths. The goal here is to understand patterns of pandemic policymaking, but from an institutional perspective, where characteristics of proposed bills, absent the text itself, reveal structural characteristics of the institutions in which these actors are acting. By exploring these patterns, whether similar or not, a deeper understanding of pandemic policymaking is possible.

In this exploration, I leverage unsupervised manifold learning to uncover the structure of policymaking on all bills related to epidemics from 1973--2020.\footnote{As addressed and justified below, and in line with \cite{waggoner2020computational}, I use ``epidemic'' and ``pandemic'' as policy heuristics interchangeably throughout, allowing for cross-temporal comparability.} To do so, I employ uniform manifold approximation and projection (UMAP) to explore pandemic policymaking, and so recover to the latent manifold of this space. As I am interested in the characteristics of the bills, rather than the text of the bill, I rely only on bill-level metadata (e.g., cosponsors, party of primary sponsor, committee assignments, etc.). My goal, then, is to understand whether characteristics of bills on similar topics are stable over time, or whether they shift in detectable ways. If stable, then this would suggest we are \textit{not} witnessing a unique period in policymaking during COVID-19, as the manifolds would map well onto each other. On the other hand, if the structure is unstable and shifting over time, then this would suggest the current era of COVID-19 policymaking is indeed \textit{unique}, relative to historical policymaking on similar issues.

Across several stages, results point to remarkable stability in the pandemic policymaking space, such that the current COVID-19 era maps extremely closely onto prior periods. Indeed, the manifolds are nearly identical, based on bill characteristics, after accounting for time. This suggests that there is less of an ``evolutionary trend'' in pandemic policymaking, where instead there is striking uniformity in this type of Congressional policymaking, despite currently operating an era of hyperpolarization \cite{gaughan2016illiberal}, deepening mass political polarization \cite{simas2020empathic}, and ineffective governance \cite{brownstein2008second}. Implications of these findings and next steps are discussed in the concluding remarks.

\section{Empirical Strategy}

As this project is interested in uncovering latent structure in a common space, but with no expected outcome, this is an unsupervised problem. Further, I assume configurations of pandemic-related bill metadata should reflect substantive patterns of \textit{aggregate} policymaking, and thus point to institutional characteristics underlying pandemic policymaking. Such an assumption, in technical terms, is that these observations (bills) lie on a common manifold, and are thus structurally related. While this manifold need not be fully connected, such that each bill to every other bill along the manifold, the expectation rather is that the bills came from a common space, which mirrors reality. Put differently, there is a common data generating process, where all legislators who author bills are acting in a common space, under common constraints, and on average, have a largely common set of goal as they author legislation \cite{waggoner2018bill}. Taken together, the unsupervised nature of the task and the assumption of a common manifold underlying these pandemic policies, the core assumption of this project is that there is latent, non-random underlying the pandemic policymaking space. The task, then, is to recover this manifold, and then compare versions of it over time, to address whether periods of pandemic policy are similar or different. As this is an exploratory project, results revealing either similarity in contours of pandemic policymaking or not, will nonetheless deepen an understanding on aggregate elite approaches to responding to major, national crises.

Importantly, the terms ``pandemic'' and ``epidemic'' often refer to instances of disease outbreak, with the former being wider spread than the latter \cite{who}. Yet, \cite{waggoner2020computational} demonstrated that when a wide array of problems, disease or otherwise, become widespread and capture national or international attention, these types of problems are often branded as ``epidemics'' by policymakers. As such, in this research I too use these terms as policy heuristics for widespread problems requiring governmental action at some level. Indeed, use of the word ``pandemic'' is largely unique to COVID-19 in the context of American policymaking, as politicians do not typically author legislation on international issues. Yet, the COVID-19 pandemic, though ultimately unparalleled in scope and impact, is related to other more commonly dubbed ``epidemics'' in America, such as the opioid epidemic or the HIV/AIDS epidemic in the 1980s, both affecting the nation and thus requiring governmental policymaking responses. As such, the loose definition of the terms found in \cite{waggoner2020computational} is based more on a policy-focused heuristic, rather than a formal public health definition. I follow the same logic in this research, and use ``pandemic'' and ``epidemic'' interchangeably to allow for a collection of a set of policies that are generally comparable over a long period of time. 

\subsection{Data}

The data were scraped from congress.gov, and include all bills related to epidemics introduced in the American Congress from 1973 to 2020 \cite{waggoner2020computational}. The data include several bill-level features: the chamber in which the bill was introduced (House or Senate), the canonical designation of the bill (e.g., resolution, joint resolution, bill, concurrent resolution)\footnote{It's important to note that naming conventions vary by chamber. For example, a \textit{resolution} might be a declaration of an idea, whereas a \textit{bill} might be more substantive.}, the Congress (two-year period) in which the bill was introduced ranging from the 93rd (1973-1974) to the 116th (2019-2020), the date of bill introduction, the long title of the bill, the party of the prime sponsor, the representing state of the prime sponsor, the representing district of the prime sponsor\footnote{$\text{dist} = 0 \forall \text{Senate}$}, number of cosponsors on the bill (non-negative integer), configuration of committees in which the bill was read\footnote{This could be a single committee or many. For example, HB 6311 in the 116th Congress, \textit{Care for COVID-19 Act}, was read in three committees (Energy and Commerce, Ways and Means, Education and Labor), whereas HB 6623, \textit{COVID-19 Language Access Act}, from the same Congress was only read in a single committee (Oversight and Reform).}, the date of last action, and finally the last action (e.g., read in committee or signed into law).

The data are split into two periods for analysis: pre-COVID (1973--2018) and COVID (2019--2020). Time is explicitly addressed in a later stage of analysis below.

For this application, I focus only on the features that can be included in the specification, which is a subset of the full feature space (e.g., raw text cannot yet be included and treated like quantitative features in UMAP applications): bill type, committee configuration for each bill, state of prime sponsor, chamber of prime sponsor, date of introduction, and date of final action. Importantly, party of the prime sponsor is omitted to allow for conditioning visualizations of the lower dimensional space on party to understand whether the lower dimensional manifold finds any discernable differences between policymaking on a partisan basis. This decision, similar to \cite{waggoner2020computational}, makes results more substantive interpretable (e.g., we can understand that Republicans may be different Democrats, compared to plots with only black dots in a two-dimensional setting). 

\subsection{Uniform Manifold Approximation and Projection}

Uniform manifold approximation and projection (UMAP) \cite{mcinnes2018umap} is a recent approach to dimension reduction, which is particularly well suited for high dimensional (e.g., $p > 4$) contexts. As in other dimension reduction approaches such as principal components analysis (PCA), the goal is to make a complex, high dimensional space more interpretable and manageable by intentionally discarding some information, but for the benefit of honing in on the most interesting variation that characterizes the data well. UMAP, though, joins this common statistical learning approach to dimension reduction with a formal mathematical foundation, based in topological data analysis (e.g., simplicial subspaces), manifold learning, and graph theory. The result is a scalable, computationally efficient (extremely fast), and mathematically grounded approach to dimension reduction. The goal of UMAP, then, is to learn a lower dimensional version of the data (i.e., low dimensional embedding, as in PCA), but assuming the data exist along a common manifold. If the manifold is recovered, a better, but more parsimonious understanding of the data can also be recovered. 

UMAP finds a lower dimensional manifold, $w_{ij}$, that captures both local structure in such a way that retains spatial relationships among observations $i$ and $j$ (via $d(i,j)$, where $d(\cdot)$ is some measure of distance) in the original high dimensional setting, $v_{ij}$. The goal is to do this, but while also retaining global structure to understand the full shape of the manifold, compared to PCA, that seeks only to \textit{maximize} variance in the raw data space to give a lower dimensional summary of that space. UMAP accomplishes retention of both global and local structure, instead, \textit{minimizing} information loss across the high ($v_{ij}$) and low ($w_{ij}$) dimensional versions of the data by optimizing a cross-entropy cost function, 

\begin{equation}
\sum_{i \neq j} v_{ij} \text{log}(\frac{v_{ij}}{w_{ij}}) + (1-v_{ij}) \text{log}(\frac{1 - v_{ij}}{1 - w_{ij}}).
\label{eq:ce}
\end{equation}

The full cost function in equation \ref{eq:ce} can be rearranged into two components, which are typically optimized by stochastic gradient descent \cite{mcinnes2018umap}, 

\begin{equation}
\sum_{i \neq j} v_{ij} \text{log}(v_{ij}) + (1-v_{ij}) \text{log}(1-v_{ij}) - v_{ij} \text{log}(w_{ij}) - (1-v_{ij}) \text{log}(1-w_{ij}).
\label{eq:c1}
\end{equation}

As such, the goal is minimize differences of local and global structure in the raw high-dimensional setting, $v_{ij}$, and the lower dimensional manifold, $w_{ij}$. 

To underscore the graph-based approach of search for an optimal manifold that underlies, and thus \textit{connects} points that lie along it (i.e., a graph), the task is to search at each point for the nearest neighbor, and then connect neighbors with an edge in a smooth way, such that the density of some region determines the decay, which moves outward from the point of interest, $i$, in a fuzzy way. In other words, higher density regions have smaller radius around each point, compared to less dense regions, which have a wide radius of decay, allowing for the manifold to connect. This satisfies the previously mentioned assumption of UMAP, which is that all points exist along a common manifold, though are not necessarily \textit{locally} connected. As such, the cost function can be re-written in graph notation as, 

\begin{equation}
\sum_{e \in G} w_h(e) \text{log}(\frac{w_h (e)}{w_l (e)}) (1-w_h(e)) \text{log}(\frac{1-w_h (e)}{1-w_l (e)}),
\label{eq:graph}
\end{equation}

where, the first term, $\sum_{e \in G} w_h(e) log(\frac{w_h (e)}{w_l (e)})$, allows for optimal recovery of the local neighborhoods, and the second term, $(1-w_h(e)) log(\frac{1-w_h (e)}{1-w_l (e)})$, allows for optimal recovery of the spacing between the local neighborhoods, thus allowing for consistent \textit{global} structure, based on consistent \textit{local} structure. 

The term ``consistent'' points to vastly important aspect of UMAP, which is one of the several improvements to t-distributed stochastic neighbor embedding (t-SNE, \cite{maaten2008visualizing}). That is, the first term in equation \ref{eq:graph}, $\sum_{e \in G} w_h(e) log(\frac{w_h (e)}{w_l (e)})$, is essentially what t-SNE is interested in doing, that is, capturing local behavior. Yet, t-SNE does so in a probabilistic way, by drawing miniature t-distributions around each point, and calculating the probability that a certain number of points \textit{should} be nearest neighbors, with some degree of uncertainty. Yet, by leaving out the second term, $(1-w_h(e)) log(\frac{1-w_h (e)}{1-w_l (e)})$, t-SNE is unable to project \textit{new} data onto the lower dimensional embedding, as the lower dimensional embedding can change at each iteration, and is thus not reproducible. 

As such, a major advantage of UMAP over t-SNE and other manifold-based dimension reduction techniques, is the ability to reproduce the same lower dimensional embedding, and thus opening up the possibility of supervised projection of new points onto the learned manifold. I take advantage of this feature in the second stage below, to directly compare pre-COVID policymaking with COVID-era policymaking. 

Further, it is important to note that while t-SNE finds nearest neighbors by calculating a series of $N$ pointwise conditional probability distributions, similarity in the context of UMAP, thus defining nearest neighbors, $v_{j | i}$, is calculated by a measure of smoothed nearest neighbor distances based on spatial proximity to each other \cite{mcinnes2018umap}, e.g.,

\begin{equation}
v_{j | i} = exp[(-d(i,j)-\rho_i)/\sigma_i]
\end{equation}

where, as before, $d(\cdot)$ is some measure of distance, not necessarily Euclidean, and $\rho$ (minimum distance to the nearest point) and $\sigma$ (some normalization factor controlling smoothness) are hyperparameters affecting smoothness of the solution \cite{mcinnes2018umap}. In practice, tuning these hyperparameters affects the tradeoff between local and global structure, in addition to several other hyperparameters discussed more in the following subsection.

\subsection{Hyperparameter Tuning}

There are five major hyperparameters that must be tuned when applying UMAP to a dimension reduction task of this sort: $k$ (the number of neighbors considered in each neighborhood search), $\rho$ (the minimum distance to the nearest point), number of epochs (number of times the algorithm sees the data), $d$ (which is the number of dimensions constraining the lower dimensional embedding; this is usually 2 in most cases, as well as in this research), and $\text{dist}$ (distance metric for pairwise distance calculations).

\begin{figure}[h!]
	\centering
	\includegraphics[scale = 0.65]{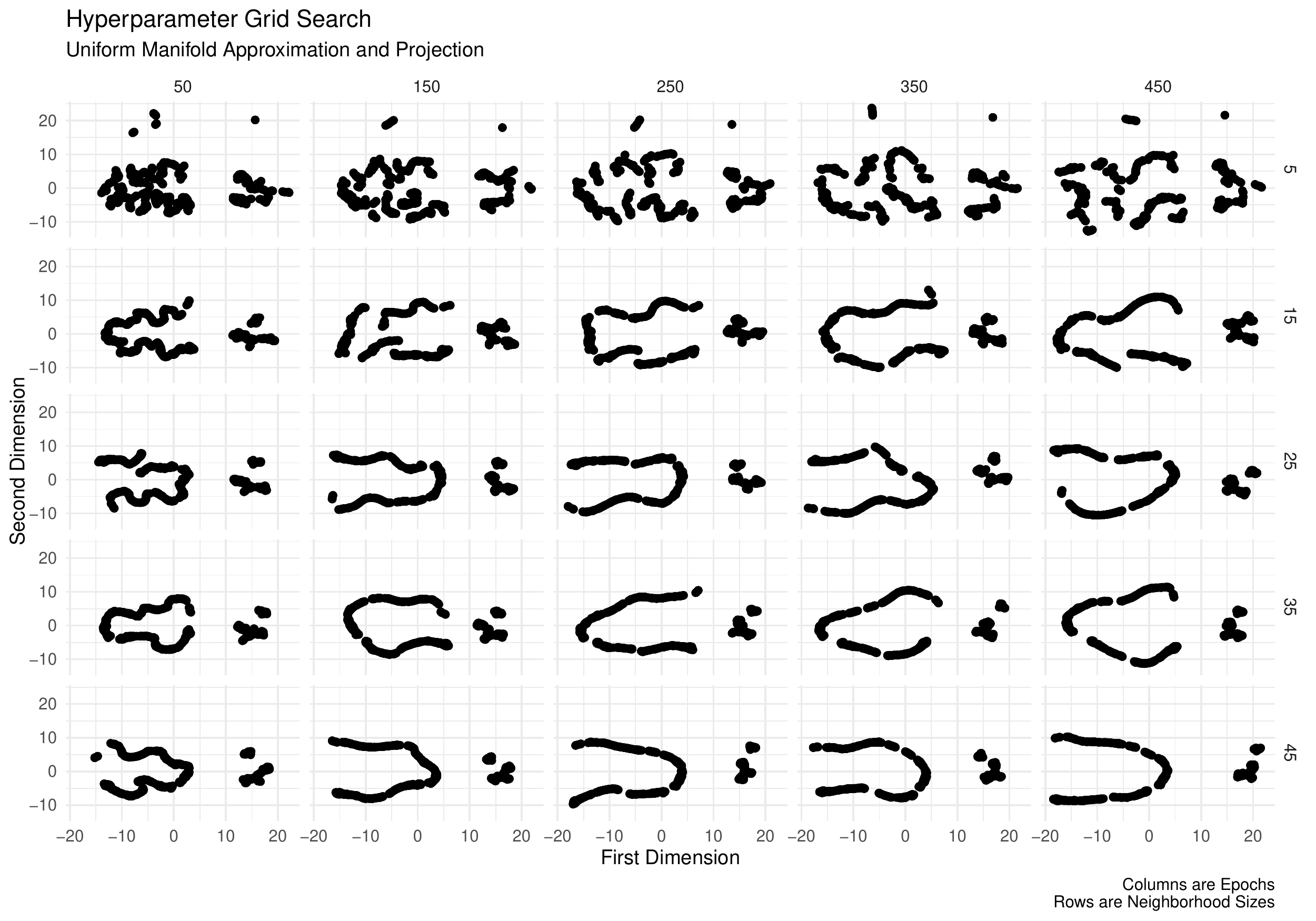}
	\caption{Grid Search of UMAP Hyperparameters}
		\label{figure:grid}
\end{figure}

As with many learning algorithms, a common approach to hone in on final values for the hyperparameters is to conduct a grid search across multiple values of each. I do so across the two main hyperparameters, $k$ neighborhood size and number of epochs, and present the results in Figure \ref{figure:grid}.

From Figure \ref{figure:grid}, a few patterns stand out. First, as expected, the more the algorithm sees the data (as the epochs increase), the clearer and more stable the patterns in the lower dimensional manifold become. Moving from left to right across the column facets in Figure \ref{figure:grid}, the separation between smaller subgroups in the data, become starker. This suggests that there is indeed some level of difference in pandemic policymaking over the full study period. These differences will be pull apart in the following analysis. 

A second notable pattern in the grid search in Figure \ref{figure:grid} is the further clarity moving from top to bottom of the figure across the row facets, showing that as the size of the neighborhood increases from 5 (top row) to 45 (bottom row), additional clarity is gained. 

Related, and substantively, as we inspect the highest values from each of these hyperparameter values in the lower right plot in Figure \ref{figure:grid}, it becomes increasingly clear that two distinct groups of pandemic-related policies take shape. Precisely what these groups are and how they are comprised is the task of the later stages of analysis below.

But, importantly, it is worth point out that the features have been scaled and standardized, meaning the axis values have no substantive meaning, as in other similar dimension reduction techniques like t-SNE. Therefore, the greatest strength of these types of manifold-based dimension reduction techniques is to visualize the reduced data space, which allows for greater insight into substantive differences that naturally exist in the higher dimensional space. Here, the first conclusion we can draw from the single Figure \ref{figure:grid}, is that there seem to be two groups of pandemic-policies characterizing this space. I transition now to pull this apart more overtly, by proceeding with hyperparameter values $k = 45$, $\text{epochs} = 450$, $d = 2$, $\text{dist} = \text{Euclidean}$, and $\rho = 0.1$.

\section{Learning the Manifold of Pandemic Policymaking}

\begin{figure}[h!]
	\centering
	\includegraphics[scale = 0.45]{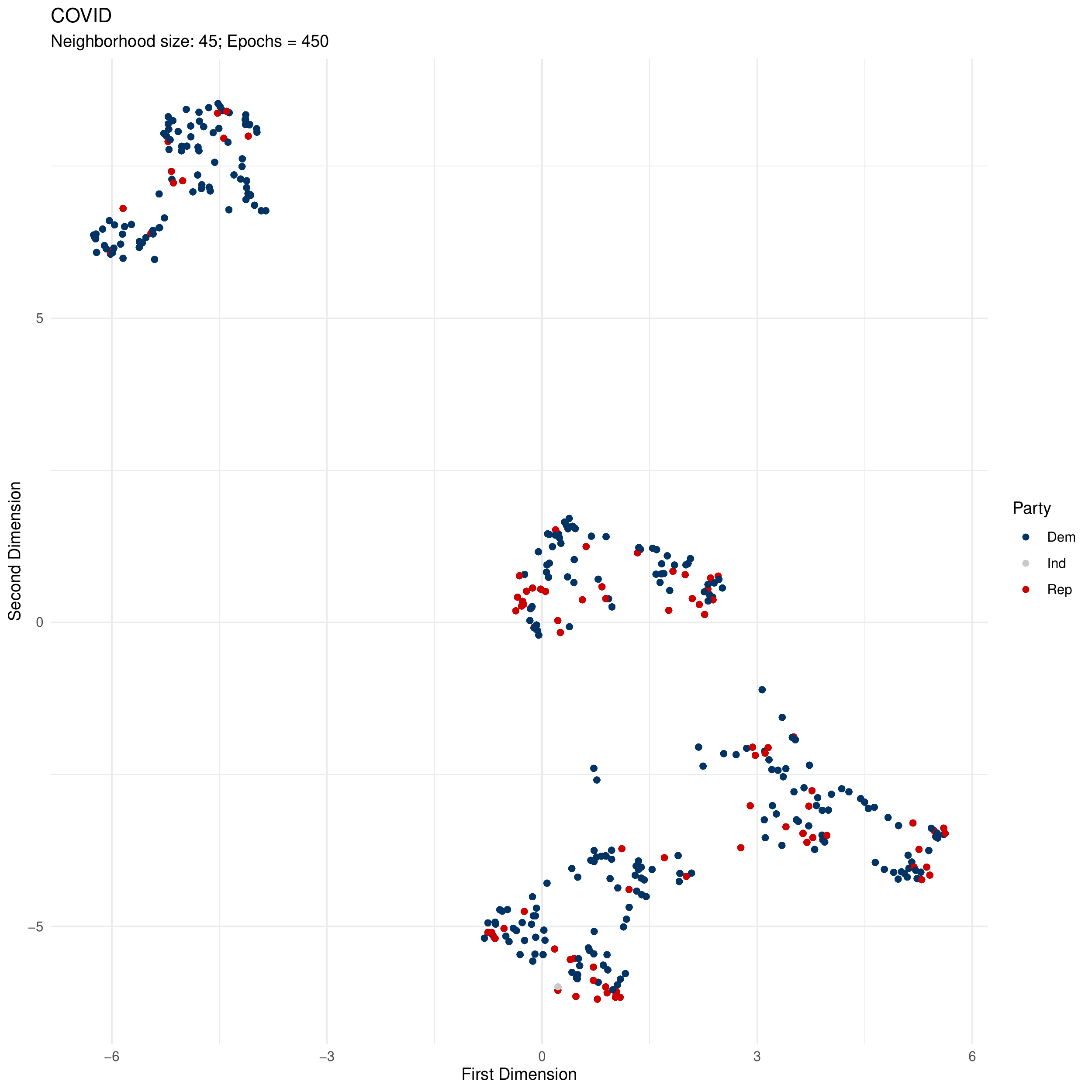}
	\caption{COVID Policymaking Structure}
	\label{figure:covid}
\end{figure}

I first present the results in Figure \ref{figure:covid} from fitting UMAP to the COVID period of policymaking. Point colors correspond with the party of the prime sponsor on the bill. This condition allows for understanding whether partisan differences exist in pandemic policymaking, in addition to the vector of additional features on which the UMAP fit is based. \\

\begin{figure}[h!]
	\centering
	\includegraphics[scale = 0.45]{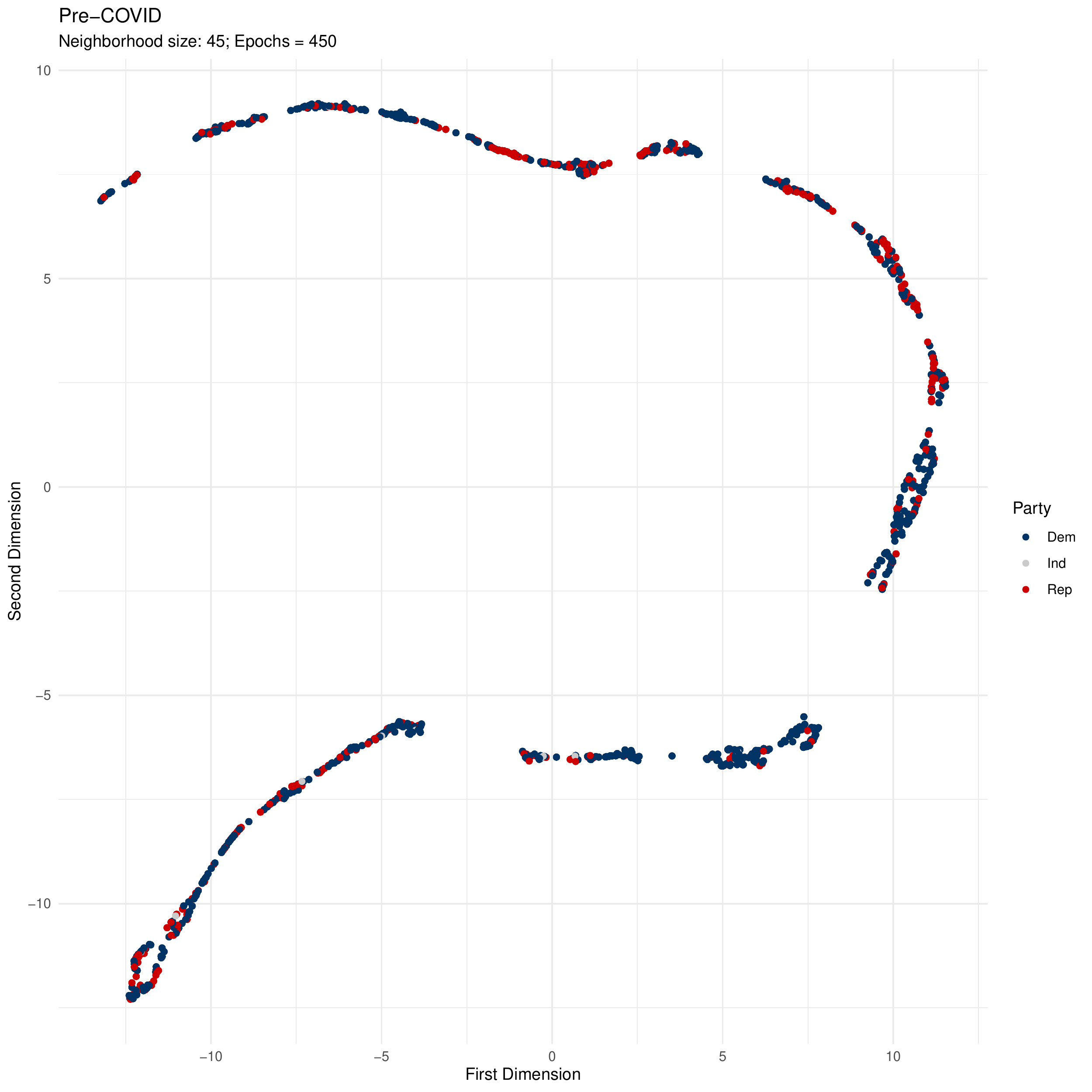}
	\caption{Pre-COVID Policymaking Structure}
	\label{figure:precovid}
\end{figure}

Building on the patterns from the grid search in Figure \ref{figure:grid}, it seems as though the distant cluster of points, apart from the ``C''-shaped cluster, is the COVID period. Again, as the key value of UMAP, and visualization of this sort is to observe where observations lie in high dimensional space, as recovered in a lower dimensional setting, the main take away from Figure \ref{figure:covid} is that there seems to be a distant pocket of Democratic (blue) bills, with only a few Republican bills in the upper left of the plot, which is distant from the remainder of the policies in the lower right of the plot. In the lower right, there is a blend of Democratic (blue) and Republican (red) bills mixed together, implying, for the most part, there are no major differences between the parties in the approach pandemic policymaking. While the distant group of mostly Democratic bills to the upper left is interesting, explicit probing of that group is beyond the scope of this analysis. Rather, I am interested in exploring whether differences exist across the two major time periods of pandemic policymaking. To this end, I turn now to present results from the UMAP fit on the pre-COVID set of policies, shown in Figure \ref{figure:precovid}. \\

Here, we can confirm that the ``C''-shaped cluster of points indeed captured the pre-COVID era of policymaking. The distribution of political parties in this cluster over a long period of time is relatively uniform, with no major partisan differences emerging. 

Of note, though, is the unique shape of the cluster, compared to the COVID period in Figure \ref{figure:covid}. It seems as though the structure is more clearly segmented, with those existing in a given cluster (e.g., the first group in the lower left of the plot) being explicitly in that cluster, compared to the rest of the space, which is also clearly and explicitly grouped. In sum, this suggests that there is some aspect of this policymaking space that clearly sorts policies into bins that do not blend for the most part. 

Stepping back and comparing trends across the COVID (Figure \ref{figure:covid}) and pre-COVID (Figure \ref{figure:precovid}) periods, the patterns in the data seem to suggest that these periods of pandemic policymaking are indeed distinct. This gives credence to the notion that we are indeed witnessing a truly unique moment in American policymaking, where political elites are engaging in policymaking in some fundamentally different way than ever before. 

\section{An Alternative Approach: Supervised Projection}

At this point on visual inspection of the two periods of pandemic policymaking, it seems as though the periods are indeed distinct in some structural, high-level way. Further, there seems to be a lack of clear partisan distinction between the periods, resulting in a lot of party overlap across both periods on average. Even still, exploration has been centered on observation and comparison of patterns in a descriptive way. 

Yet, recall one of the prime benefits of UMAP is to learn a lower dimensional representation of the data that retains and balances global \textit{and} local structure in a reproducible way given the lack of reliance on a probabilistic neighbor search as in t-SNE. In practice, this feature of a reproducible solution from a UMAP fit allows for \textit{supervised} dimension reduction, where new data can be projected onto the learned manifold, and the two are able to be compared directly. If there is stability in the structure across the sets of data, then the learned manifold is likely capturing true, unique features of the data, such that new, unseen data can be mapped closely onto that manifold. Yet, if there is a lack of structure in the space, then the learned manifold will be different from the manifold found when mapping new data to it. 

For current purposes, this benefit of UMAP helps with a more direct comparison of the two eras, beyond simply visually exploring two fits of UMAP as has been done to this point. As such, in this section, a fit UMAP on the pre-COVID era and learn the manifold, and then project the COVID era data onto the learned manifold from the pre-COVID era. In so doing, I treat the COVID era as ``new'' or ``testing'' data, and the pre-COVID data as ``training'' data.\footnote{Though this strategy is substantively motivated for present purposes, as a check on these findings, I also randomly split the data in training and testing sets and proceed with the same supervised task in the following subsection and discuss results, which are strongly supportive of patterns found at this stage.}

\begin{figure}[h!]
	\centering
	\includegraphics[scale = 0.45]{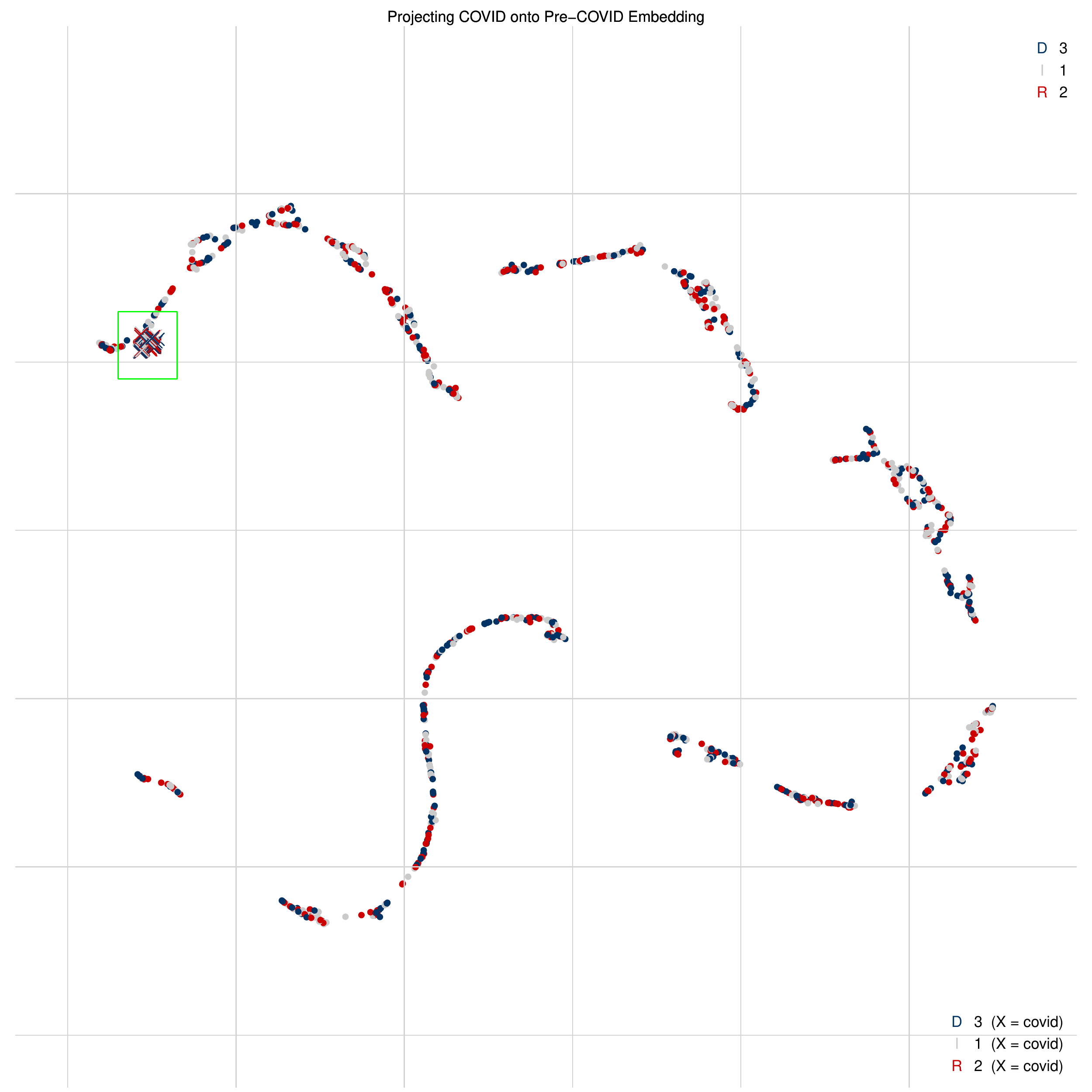}
	\caption{Projecting COVID onto Pre-COVID Period}
	\label{figure:projno}
\end{figure}

First, I present the UMAP results on the full feature space in Figure \ref{figure:projno}. Here, all features, including time-dependent features (e.g., date of introduction, Congressional period), are included. The circle points in the figure represent the learned manifold from the training/pre-COVID set. The X's represent the projection of the test/COVID period. 

This view of the two periods is in line with findings to this point, where the COVID period (denoted by the green box) seems to be a very unique period of policymaking different from nearly every other period in pandemic policymaking. This is seen by the tiny position of COVID onto the full manifold, suggesting it projects very poorly onto this space.

If the COVID period were truly different from the pre-COVID period, then we might expect to see the COVID period project poorly onto the manifold from the pre-COVID period, e.g., a cluster of points distant from the main cluster of the pre-COVID period, which would be different from the pattern in the grid search in Figure \ref{figure:grid}, as the grid search was \textit{not} based on projection. Yet, we see the COVID period very clearly and precisely projecting onto a very small, but specific part of the pre-COVID manifold. While not spread across the full pre-COVID period, the COVID period nonetheless clearly maps onto a the lower dimensional manifold. Given the previous pattern in Figure \ref{figure:precovid} of precise, but still distinct clusters of policies very clearly spread out in the same space, the inclusion of time-dependent features is likely influencing the shape of the manifold. This is likely the case, because, for example, bills sponsored in 1980 will be treated very differently than bills sponsored in 2020, given their different values along this feature. The same is true for other time-dependent features, like the Congressional period feature, which is treated as a non-negative integer (e.g., ranging from 93-116).

To explore whether this is the case, I turn now to fit a new version of UMAP on a subset of the feature space, \textit{excluding} all time-dependent features. We are left with five features, still withholding party affiliation to allow for conditioning the points on the plot: bill type, cosponsors, committee configurations, state of primary sponsor, and the chamber of the primary sponsor. Though significantly pairing down the space, we are still left with a high dimensional problem, as we are consider a five-dimensional feature space. 

\begin{figure}[h!]
	\centering
	\includegraphics[scale = 0.45]{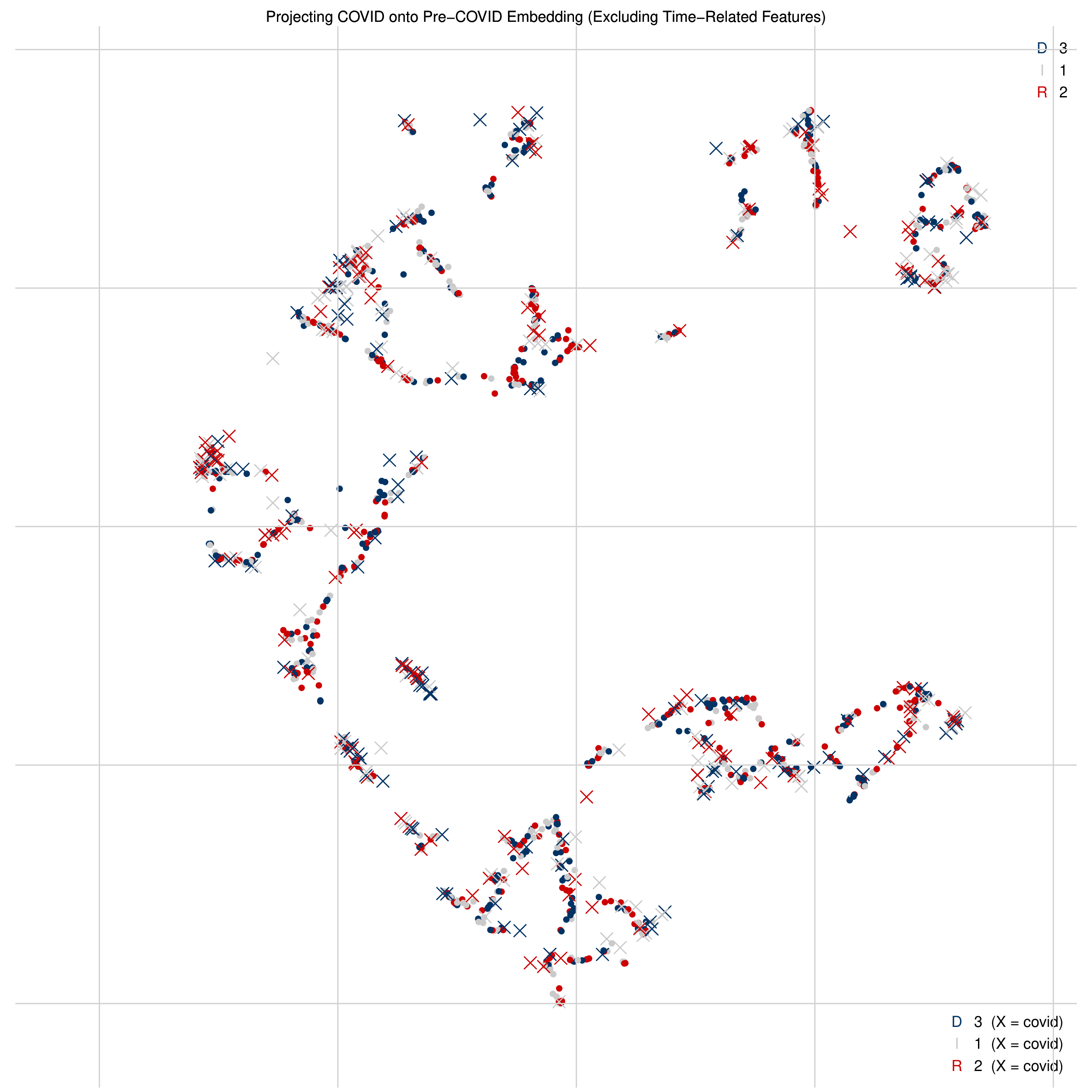}
	\caption{Projecting COVID onto Pre-COVID Period: Addressing Time-Related Features}
	\label{figure:proj}
\end{figure}

As before, upon learning the lower dimensional manifold of the pre-COVID period (minus time-dependent features), I project the COVID period onto the manifold to explore whether and to what degree the manifolds align. Results are presented in Figure \ref{figure:proj}, where again, circle points are pre-COVID bills, X's are COVID bills projected onto the space, and colors vary by party of the prime bill sponsor.

We now get a very different view of the pandemic policymaking space. Now, considering the role of time by dropping the time-dependent features from the fit, the patterns of pre-COVID and COVID map extremely closely on each other. Substantively, this suggests that the two periods, spanning a near 50-year period of Congressional policymaking, look very similar to each other. Further, the structure shrinks as well, with the ``C''-shape of the manifold disappearing, and the precise grouping (presumably by time) from the earlier patterns in Figures \ref{figure:precovid} and \ref{figure:projno}, now disappears. Indeed, if the COVID period of policymaking were truly unique, we might expect poor projection and no evidence of clear mapping across the periods. Yet, as clearly shown in Figure \ref{figure:proj}, the configuration of pandemic-sponsoring legislators look quite similar to each other over time. Importantly, from these results, as they are exploratory, it is impossible to say whether the patterns we see today \textit{are a function} of the historical approach to pandemic policymaking. Target causal methods would be required for that conclusion. But what is very clear from these results is that not much has changed over 50 years regarding how legislators approach pandemic policymaking.

\subsection{Random Data Splitting for Supervised Projection}

\begin{figure}[h!]
	\centering
	\includegraphics[scale = 0.45]{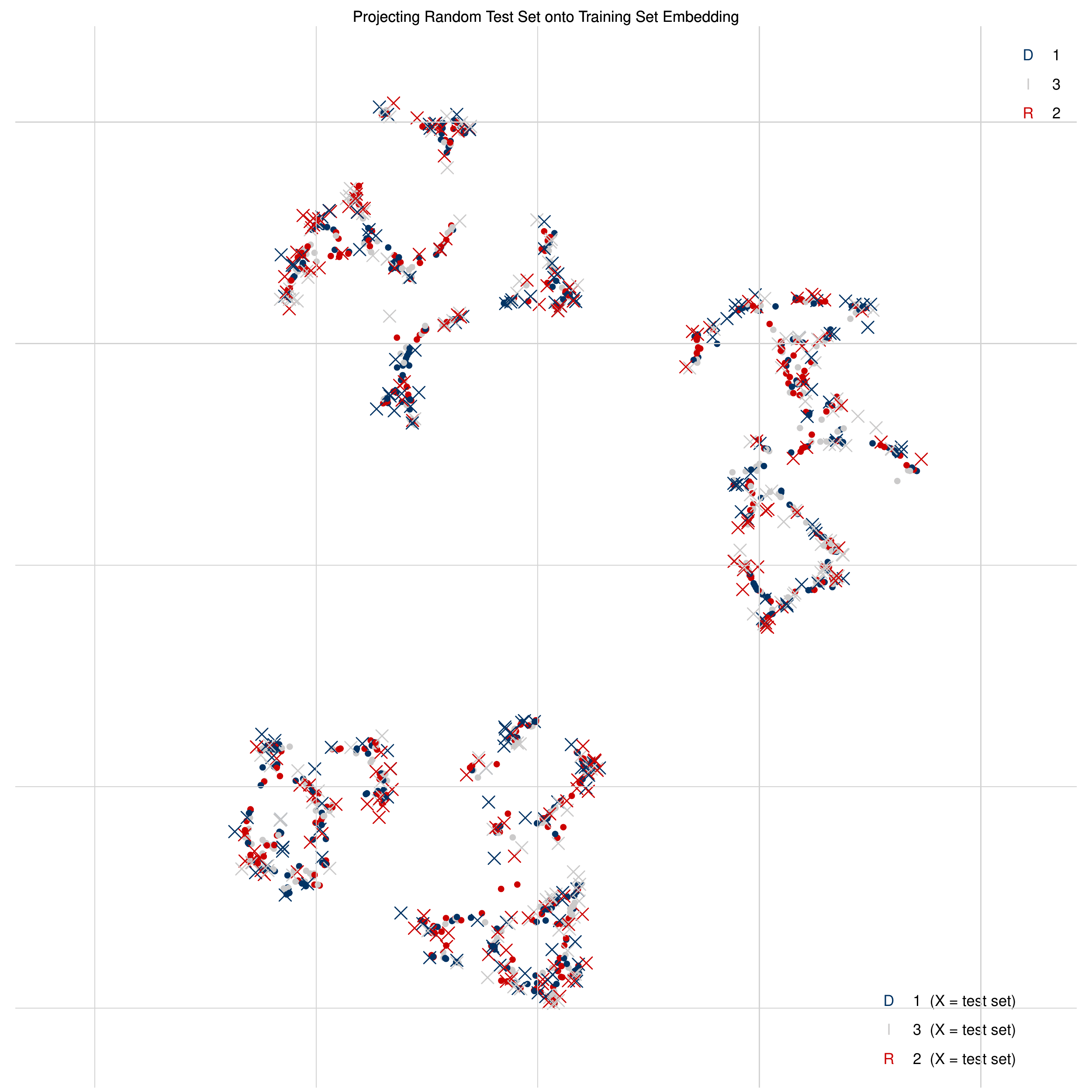}
	\caption{Pandemic Policymaking Structure with Random Data Split}
	\label{figure:tt}
\end{figure}

Even though the choice of setting the pre-COVID era as the training set and the COVID era as the testing set was substantive motivated, to directly compare the patterns of pandemic policymaking across the two periods, I offer a final check on these patterns in this section. To do so, and thus validate the patterns found thus far, particularly those in Figure \ref{figure:proj}, I randomly split the full data space into training and testing sets, regardless of whether policies were in the COVID or pre-COVID periods. Note, I retain the same proportion of bills in each major period for direct comparison ($\approx 0.66$ in the training/pre-COVID set and $\approx 1 - 0.66$ in the testing/COVID set). Further, building on the findings showing striking similarity across the periods in Figure \ref{figure:proj}, here I also exclude time-specific features (e.g., Congress period, date of introduction). Results are presented in Figure \ref{figure:tt}.

In Figure \ref{figure:tt}, as before in Figure \ref{figure:proj}, the circles represent the manifold learned on the training set, and the X's represent the projection from the testing set. Also, colors are still conditioning on party of the prime bill sponsor. The key take away is that the patterns are extremely similar, with the projection of a random set of test points mapping strikingly closely onto the manifold learned from the training set. This strengthens the findings and conclusions drawn from the previous section that, indeed, the two periods of pandemic policymaking are virtually indistinguishable, as it relates to partisan patterns, committee configurations of bills, cosponsorship, and so on for all features included in the UMAP procedure. Ultimately, this suggests that pandemic policymaking, though we are in a current period of hyperpolarization and political division, is a relatively stable phenomenon. This conclusion, though made on the basis of an exploratory study, provides a glimmer of  hope in a current American political climate marked by bitter division and dislike for members of opposing political parties, such that when crisis strikes, politicians seem to have an informal code of policymaking, which seems to be closely adhered to over a near 50 year period of pandemic policymaking.

\section{Concluding Remarks}

In this paper, I set out to explore and uncover the lower dimensional manifold of American policymaking related to ``epidemics'' broadly defined. In so doing, a second order goal was to understand whether the unique institutional context of political division and polarization also influence the types of policies aimed at addressing COVID-19. By focusing on COVID-19 and institutional structure, the concept of pandemic policymaking places COVID-19 legislation into historical context. The result is a comparison of patterns of policymaking over time, to explore whether past approaches to epidemic-related policymaking map onto the current approach of policymaking on COVID-19. 

Overall, though the initial stages of the exploration seemed to suggest the current era of COVID-19 pandemic policymaking is distinct from prior periods, a more targeted exploration showed that once explicitly accounting for time, COVID-19 policymaking mapped extremely closely onto prior periods of pandemic and epidemic-related policymaking. This suggests, then, that though the political and institutional contexts have changed, becoming increasingly bitter and divided, the approaches to and patterns of pandemic policymaking have remained largely stable over time. The substantive conclusion, then, is that the current era of COVID-19 policymaking looks very similar to prior eras of policymaking on a host of epidemics, all at varying scales, implying a degree of (restored) hope in the prime institution for policymaking in America. Indeed, a more stable, and perhaps even formulaic approach seems to characterize Congressional pandemic policymaking, regardless of the surrounding political and institutional context as well as the nature and scope of the epidemic(s) in question. These patterns were corroborated using random data splitting and then replicating the supervised projection task. Results remained strikingly similar across all periods, regardless of the cases ending up in the training or testing test, thereby strengthening conclusions of uniformity in pandemic policymaking. 

Future work might pick up on these results by considering either different periods of political history, different substantive topics (e.g., the economy, elections, and so on), or different governments around the world beyond the American case. 

On the technical side, future work might pick up on the methodological approach (UMAP), but relaxing the strong assumption that all observations exist along a common manifold, though sparse connections are still possible. A common manifold simply may not capture reality, where different data generating processes may result in actors in a common space acting fundamentally differently from others. Such an extension of UMAP would be akin to anomaly detection, but in the context of, e.g., deriving \textit{multiple} manifolds from a common space to explain such sparsity.

\clearpage

\bibliographystyle{IEEEtran}
\bibliography{pp}

\end{document}